\documentstyle[aps,preprint,amssymb,floats,graphicx]{revtex}

\newcommand{\Pomeron }{\mathbb{P}}        
\newcommand{\T}[1]{{\bf #1}_{T}}          

\newcommand{\LBF}[1]{\mbox{\large\bf #1}}

\begin{document}
\draft
\tighten

\preprint{
        \parbox{1.5in}{%
           \noindent
           PSU/TH/189 \\
           hep-ph/9709499
        }
}

\title{Proof of Factorization for Diffractive Hard Scattering}

\author{John C. Collins\footnote{
    Electronic address: {\tt collins@phys.psu.edu}.
}}
\address{Penn State University,
        104 Davey Lab, University Park PA 16802, U.S.A.
}

\date{September 6, 1999}

\maketitle
\begin{abstract}%
    A proof is given that hard-scattering factorization is valid
    for deep-inelastic processes which are diffractive or which
    have some other condition imposed on the final state in the
    target fragmentation region.
\end{abstract}


\section{Introduction}
\label{sec:introduction}

In this paper, I show how to prove hard-scattering factorization
for diffractive deep-inelastic processes, and certain related
processes.  This is an important topic because it is known
\cite{oldDY,CW,DEL,CFS} that factorization fails for
hard processes in diffractive hadron-hadron scattering (like
diffractive Drell-Yan).\footnote{
    Note that this knowledge predates QCD. Within the context of
    pre-QCD parton-model ideas it was shown that there are
    factorization-breaking terms \cite{oldDY} in both the
    diffractive and non-diffractive parts of the Drell-Yan
    process, and that these terms cancel \cite{CW,DEL} in the
    inclusive cross section, which is the sum of the diffractive
    and non-diffractive parts. This result forms part of the
    proof of factorization for inclusive hard processes in QCD
    \cite{fact-proof}.
}
Moreover, the violation of factorization appears to be confirmed
by experiment \cite{CDF,D0,ACTW,goulianos}.  So we must determine
those diffractive processes, if any, for which factorization is
actually predicted by QCD.

The precise form of the factorization property that I prove has
been stated by Kunszt and Stirling \cite{KS}, and by Berera and
Soper \cite{BS1,BS2}, as a full QCD generalization of the
Ingelman-Schlein model \cite{IS}, but shorn of the Regge
hypotheses.  It is the same as factorization for inclusive hard
processes, except that parton densities are replaced by
diffractive parton densities. We can say that Ingelman-Schlein
\cite{IS} factorization is hard-scattering factorization, such as
is proved in the present paper, together with Regge factorization
for the pomeron exchange.

I will prove the theorem not only for diffractive deep-inelastic
processes, but for any deep-inelastic process where a requirement
is imposed on the final state in the target fragmentation region.
Any requirement that is fixed relative to the beam is allowed:
e.g., that there be detected particle(s) of particular kind(s)
carrying some particular fraction of the beam's momentum and
carrying some given transverse momentum.  Hence the proof applies
to the fracture function formalism of Trentadue and Veneziano
\cite{TV}, for deep-inelastic processes\footnote{
    Note that since factorization fails for diffractive hard
    processes in hadron-hadron scattering, it follows that the
    fracture function formalism also fails in hadron-hadron
    scattering.  The proof given by Trentadue and Veneziano
    does not treat the soft exchanges which break factorization
    in hadron-hadron scattering.
}.
Factorization for diffractive scattering is a special case
\cite{GV} of fracture function factorization.

Furthermore, it is possible to discuss any of the normal hard
scattering processes which are lepton induced: in addition to the
deep-inelastic cross section itself, the proof applies, for
example, to the case where jets of large transverse momentum are
detected and where particular particles in the `current
fragmentation region' are detected.

The proof in the present paper justifies, from fundamental
principles, the analysis \cite{H1,ZEUS} of diffractive
deep-inelastic processes in terms of parton densities in the
pomeron. Note that the only real use of the pomeron in these
analyses is as a label for a particular power law for the $x_{\Pomeron }$
dependence of diffractive cross sections, with the exponent
actually being a free power. Indeed, the QCD analysis by H1
\cite{H1}, which has two phenomenological power
laws, is also covered by the theorem proved in this paper.
However, I will not at all address the separate and important
question of whether {\em Regge} factorization is also valid.
Regge factorization relates, for example,  the power of $x_{\Pomeron }$
measured in diffractive deep-inelastic scattering to the power of
$s$ measured in hadron-hadron elastic scattering.

Berera and Soper \cite{BS2} provided arguments that
hard-scattering factorization should be true in diffractive
lepton-induced processes, and the present paper
completes the proof. The bulk of the proof follows the usual
methods \cite{fact-proof,CS} for proving factorization, and, as
pointed out by Berera and Soper \cite{BS2}, the only new element
that is needed is a proper treatment of the soft-gluon
cancellation for the processes in question. The essential point
of the present paper is to show that there exists a contour
deformation that permits the methods of Collins and Sterman
\cite{CS} to be used.

\section{Factorization, parton densities}
\label{sec:factn}

In this section, I will review the factorization theorem that is
to be proved.

As stated above, the factorization theorem for diffractive hard
processes has the same form as for inclusive processes.  For
example, for diffractive deep-inelastic scattering
$e + p \to  e' + X + p'$, we have\footnote{
    For the purposes of this paper, I define $F_{2}^{D}$ to be the
    value of $F_{2}$ computed from those events containing a
    final-state proton $p'$ with the specified kinematics.  So
    the use of the word `diffractive' to describe the process is
    not really correct. Our definition is the one used by the H1
    experiment \cite{H1}, and it contrasts with the definition
    used by the ZEUS experiment \cite{ZEUS}, which subtracts the
    non-diffractive contribution. Of course, given the
    `diffractive' $F_{2}^{D}$ defined here, one can extract the leading
    power at small $x_{\Pomeron }$, which, at least for our present
    purposes, is the definition of the truly diffractive part.
    Factorization for the complete $F_{2}^{D}$, as defined here,
    implies factorization for the purely diffractive part, with
    the diffractive parton densities $f_{i}^{D}$ being replaced by
    their diffractive components.
}
\begin{equation}
   F_{2}^{(D)}(x_{\rm bj}, Q, x_{\Pomeron }, t) =
   \sum _{i} C_{2i} \otimes f_{i}^{D}
   + \mbox{non-leading power of $Q$} .
\label{theorem}
\end{equation}
Here, $x_{\rm bj}$ and $Q$ are the usual deep-inelastic variables,
$x_{\Pomeron }=1-q\cdot p'/q\cdot p$ is the fractional loss of longitudinal
momentum by the diffracted proton\footnote{
    Of course, the proton may be replaced by any other hadronic
    state, e.g., a nucleus.
}, and $t=(p-p')^{2}$ is the
invariant momentum transfer from the diffracted proton, while
$\otimes$ signifies a convolution of the hard-scattering
coefficient $C_{2i}$ with the diffractive parton density $f_{i}^{D}$. The
factorization theorem applies when $Q$ is made large while $x_{\rm
bj}$, $x_{\Pomeron }$, and  $t$ are held fixed. It asserts not only that an
expansion of the form of Eq.\ (\ref{theorem}) is true, but also
that
\begin{itemize}

\item
    $C_{2i}$ is the {\em same} hard scattering coefficient as in
    ordinary (inclusive) deep-inelastic scattering (DIS), with
    $i$ being a label for parton flavor (gluon, $u$-quark, etc.).

\item
    The diffractive parton densities $f_{i}^{D}$ are those defined by
    Berera and Soper \cite{BS2}, as suitable `cut matrix
    elements' of the same operators that define ordinary parton
    densities.

\item
    They therefore obey exactly the same DGLAP evolution
    equations as ordinary parton densities.

\end{itemize}

Generalizations of the theorem that are covered by the proof in
this paper are of two kinds:
\begin{itemize}

\item
    The requirement that there be a diffracted proton $p'$ in the
    final-state may be replaced by any other requirement in the
    `target fragmentation region' that is fixed relative to the
    initial hadron.  For example, $p'$ may be a neutron, or it
    may be replaced by a two-pion state of some invariant mass
    that has a fraction $1-x_{\Pomeron }$ of the longitudinal momentum of
    $p$ and that has some given value of $t$. (Longitudinal
    momentum must be interpreted in the sense of the appropriate
    light-cone momentum, so that the definition of the parton
    densities is invariant under longitudinal boosts.)

\item
    Any other standard hard process may be considered.  Then the
    coefficient $C_{2i}$ is replaced by the appropriate coefficient
    for the process, times fragmentation functions if necessary.
    Thus the theorem applies to the longitudinal structure
    function $F_{L}^{D}$, and to differential cross sections for jet
    production in the `current fragmentation region'.

\end{itemize}

The first generalization implies that the theorem applies at all
$x_{\Pomeron }$ away from zero, and not just to the diffractive region of
small $x_{\Pomeron }$. This justifies the analysis \cite{H1} given by
H1, who analyzed $F_{2}^{D}$ in terms of two powers of $x_{\Pomeron }$, both a
leading diffractive power, and a non-leading power. It also
justifies the fracture function formalism of Trentadue and
Veneziano \cite{TV}, but only for deep-inelastic processes. Note
that Trentadue and Veneziano define their cross sections to be
integrated over the transverse momentum of the final-state hadron
$p'$.  This complicates the formalism:  Whereas the diffractive
parton densities without the integral over transverse momentum
obey standard DGLAP evolution equations, the corresponding
equations for fracture functions in \cite{TV} are more
complicated, since the outgoing particle $p'$ may be at large
transverse momentum and thus be associated with the hard
scattering.  The theorem proved here does not need the integral
over the transverse momentum of $p'$.

\section{Perturbative proof}
\label{sec:P.Proof}

As was explained by Berera and Soper \cite{BS2}, the proof of
factorization for diffractive hard processes is the same as for
inclusive hard processes \cite{fact-proof}, except for the
treatment of the cancellation of soft exchanges.

\subsection{Regions}
\label{sec:regions}

\begin{figure}
    \begin{center}
        \includegraphics[scale=0.5]{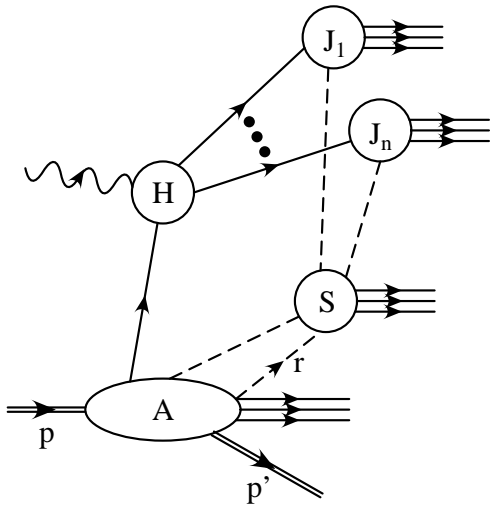}
    \end{center}
\caption{Leading regions for diffractive DIS.}
\label{fig:regions}
\end{figure}

The leading regions of Feynman graphs for amplitudes for
diffractive deep-inelastic scattering may be represented as in
Fig.\ \ref{fig:regions}, the analysis \cite{LS} being
independent of the diffractive requirement. There is a subgraph
$A$ consisting of lines collinear to $p$ and $p'$.  One
parton\footnote{
    Plus arbitrarily many gluons with scalar
    polarizations\cite{Extra.Gluons}, if we
    are in a covariant gauge.  These gluons are a gauge artifact.
}
from $A$ is incident on the hard subgraph $H$, consisting of
lines of virtuality of order $Q^{2}$ connected to the virtual
photon.  From $H$ are produced one or more lines that go into jet
subgraphs, $J_{1}$, $\cdots$.  There may be a soft subgraph $S$ (not
necessarily connected) consisting of low momentum lines (in the
Breit frame); it is joined by {\em gluon} lines to the `jet
subgraphs' $A$ and $J_{i}$.  Some lines from $S$ may go into the
final state.

It is important that it is only necessary to consider regions
where the contours of the integrations over loop momenta are
trapped. To define a unique contribution from each region of the
form of Fig.\ \ref{fig:regions}, subtractions should be devised
to avoid double counting from the different regions that
contribute for a single graph.  This issue is the same as for
non-diffractive scattering, so we do not need to treat it here.

To analyze the process quantitatively, we use light-cone
coordinates\footnote{
    $V^{\mu } = (V^{+}, V^{-}, \T{V})$, where $V^{\pm }=(V^{0}\pm V^{z})/\sqrt
2$.
}
in the Breit frame so that
\begin{eqnarray}
   q^{\mu } &=& \left( -\frac {Q}{\sqrt 2}, \frac {Q}{\sqrt 2}, \T0 \right),
\nonumber\\
   p^{\mu } &=& \left( \frac {Q}{x_{\rm bj}\sqrt 2},
                 \frac {m^{2}x_{\rm bj}}{Q\sqrt 2},
                 \T0
           \right) ,
\nonumber\\
   {p'}^{\mu } &=& \left( \frac {(1-x_{\Pomeron })Q}{x_{\rm bj}\sqrt 2},
                    \frac {(m^{2}+\T p^{2})x_{\rm bj}}{Q\sqrt 2(1-x_{\Pomeron
})},
                    \T p
             \right) .
\end{eqnarray}

\subsection{Single soft gluon attaching to jet}
\label{sec:single.soft}

By definition a soft momentum $k^{\mu }$ is one all of whose components
are much less than $Q$ in the Breit frame: $|k^{\mu }| \ll Q$.

\begin{figure}
    \begin{center}
        \includegraphics[scale=0.5]{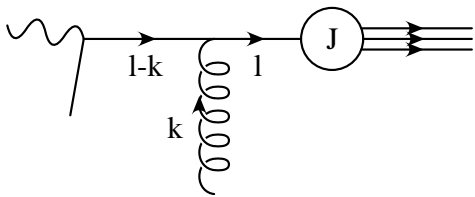}
    \end{center}
\caption{Soft gluon attaching to jet.}
\label{fig:gluon.to.jet}
\end{figure}

As a first example, which is readily generalized, let the hard
scattering be the Born graph and let a soft gluon of momentum
$k^{\mu }$ attach to the outgoing quark (Fig.\ \ref{fig:gluon.to.jet}).
We will show that after a suitable approximation in the jet
subgraph, a Ward identity can be applied to factor out the soft
attachment.  The relevant factor in the jet subgraph is
\begin{equation}
   J^{\mu }(l,k) = \frac {1}{(l-k)^{2}-m^{2}+i\epsilon } \Gamma ^{\mu } ,
\label{soft.gluon}
\end{equation}
where $m$ is the quark mass, and $\Gamma ^{\mu }$ is the vertex which couples
the gluon to the jet subgraph, together with the
attached numerator factors.  The jet momentum $l^{\mu }$ is
$(0,Q/\sqrt 2,\T0)$, plus terms that are smaller by a power of $Q$.
The largest component of $\Gamma ^{\mu }$ is $\Gamma ^{-}$ (by a power of $Q$),
so it
is a good approximation to replace $\Gamma ^{\mu }$ by $\Gamma ^{-}n_{J}^{\mu
}$, where
$(n_{J}^{+},n_{J}^{-},{\T n}_{J}) = (0,1,\T0)$.

Suppose first that all components of $k^{\mu }$ are comparable in size.
Then it is a good approximation to replace $(l-k)^{2}-m^{2}+i\epsilon $ by
$l^{2}-m^{2}-2l^{-}k^{+}+i\epsilon $, that is, to replace $k^{\mu }$ by its $+$
component
everywhere in $J$.  Thus
\begin{eqnarray}
    J^{\mu }(l,k) &=& J^{-} \LBF( l, (k^{+},0,\T0) \LBF) n_{J}^{\mu }
                + \mbox{power correction}
\nonumber\\
            &=& k^{+}J^{-} \LBF( l, (k^{+},0,\T0) \LBF)
                \frac {n_{J}^{\mu }}{k^{+}}
                + \mbox{power correction} .
\label{soft.approx}
\end{eqnarray}
The $k^{+}J^{-}$ factor is of a form to which a Ward identity can be
applied: a Green function of the gluon field contracted with the
gluon's momentum.
If Eq.\ (\ref{soft.approx}) is correct, then we can apply the
argument used by Collins and Sterman in the proof of
factorization for inclusive $e^{+}e^{-}$-annihilation \cite{CS}, and
factorization would be true for our process also.

To derive Eq.\ (\ref{soft.approx}), we assumed that all
components of $k^{\mu }$ are comparable, so that the largest term in
$k^{2}-2J\cdot k$ is $-2J^{-}k^{+}$.  The argument fails
if $k^{+}$ is too small compared to the other components of $k$.
Exactly the same problem had to be overcome in the proofs of
factorization for inclusive $e^{+}e^{-}$-annihilation \cite{CS} and for
the Drell-Yan cross section \cite{fact-proof}, etc.

Now, in the dangerous region $|k^{+}k^{-}| \ll \T k^{2}$,\footnote{
    This region was called the Coulomb region in \cite{CS}.  Note
    that if $k^{+}$ is smaller than the other components of $k$ but
    $|k^{+}k^{-}|$ is comparable with $\T k^{2}$, then $|k^{-}/k^{+}| \gg 1$,
so
    that $k$ is collinear to $J$ rather than being in the soft
    region, which is our present concern.
}
so that the only nearby pole in $k^{+}$ is the explicit pole in Eq.\
(\ref{soft.gluon}).  We may therefore deform the $k^{+}$ integration
contour away from the pole and out of the dangerous region.  This
is exactly the same argument used for $e^{+}e^{-}$ annihilation by
Collins and Sterman \cite{CS}. We must interpret the $n_{J}^{\mu }/k^{+}$
factor in Eq.\ (\ref{soft.approx}) as $n_{J}^{\mu }/(k^{+}-i\epsilon )$, so
that the
pole at $k^{+}=0$ does not interfere with the contour deformation.

\begin{figure}
    \begin{center}
        \includegraphics[scale=0.5]{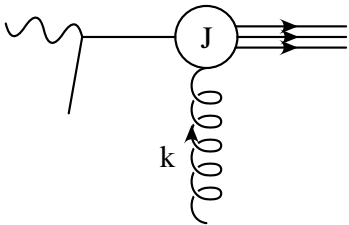}
    \end{center}
\caption{Soft gluon attaching anywhere to jet subgraph.}
\label{fig:gluon.to.jet.all}
\end{figure}

\begin{figure}
    \begin{center}
        \includegraphics[scale=0.5]{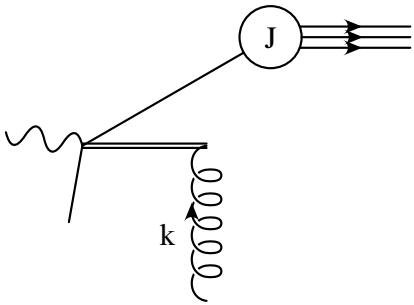}
    \end{center}
\caption{Result of summing over all graphs of the form of Fig.\
    \protect\ref{fig:gluon.to.jet.all}.}
\label{fig:gluon.to.jet.sum}
\end{figure}

The soft approximation Eq.\ (\ref{soft.approx}) therefore applies
over the whole of the soft region for $k$, on the deformed
contour.

Exactly the same contour deformation and the same approximation
can be applied to all attachments of the soft gluon to a
final-state jet subgraph, Fig.\ \ref{fig:gluon.to.jet.all}.  The
reason is that \cite{CS}, just as in $e^{+}e^{-}$-annihilation, all
interactions of soft gluons with the jet are in the final state
relative to the hard scattering. Because the direction of the
contour deformation is the same in all cases, Ward
identities\footnote{
    To implement the Ward identities correctly, account must be
    taken of graphs where the hard scattering is coupled to the
    jet subgraph $J$ by extra gluons of scalar polarization as
    well as the explicitly written quark line.  This part of the
    argument is identical to the same part of the argument for
    inclusive scattering, and so we do not need to go into the
    details.
}
can be applied consistently to factor the soft gluon out of the
jet subgraph.  The result is shown in Fig.\
\ref{fig:gluon.to.jet.sum}, where the double line represents an
eikonalized quark propagator, $1/(k^{+}-i\epsilon )$.

\subsection{General soft gluon attachments to jet subgraphs}
\label{sec:general.soft.to.J}

The argument in the previous section, \ref{sec:single.soft}
immediately generalizes, exactly as in the proof \cite{CS} of
factorization for inclusive $e^{+}e^{-}$-annihilation, to any
attachments of the soft subgraph to any of the final-state jets
subgraphs in Fig.\ \ref{fig:regions}.  Provided that we can also
apply the argument to soft-gluon attachments to the $A$ subgraph,
a sum over real and virtual emission of soft gluons can be used,
just as in $e^{+}e^{-}$-annihilation, to cancel the complete soft gluon
factors.  The cancellation only concerns a kinematic region
unaffected by the diffractive requirement on the final state.

As explained out by Berera and Soper \cite{BS2}, the desired
factorization theorem immediately follows.

\subsection{Single soft attachment to $A$}

However, we cannot apply the same argument to the attachment of
a soft gluon to the $A$ subgraph, since this subgraph contains
both initial- and final-state interactions.  The graph of a
typical leading region, Fig.\ \ref{fig:regions} illustrates this.
We have labeled one of the soft gluons attaching to $A$ by its
momentum $r$. The appropriate soft approximation is
\begin{equation}
   A^{\mu }(r, p, \dots) =
   r^{-}A^{+}
   \LBF( (0,r^{-},\T0), p, \dots \LBF)
   \frac {n_{A}^{\mu }}{r^{-}} \, + \, \mbox{power correction} ,
\label{soft.approx.A}
\end{equation}
where $n_{A}^{\mu } = (1,0,\T0)$. This approximation is valid only if $r^{-}$
is not too small.  The obvious generalization of the argument in
Sect.\ \ref{sec:single.soft} would have us deform $r^{-}$ away from
the poles of denominators in $A$ to avoid the region where the
approximation fails.  Precisely because $A$ contains both
initial- and final-state interactions, there are nearby poles in
both the upper and lower half-planes, and we cannot deform the
$r^{-}$ contour to where the soft approximation is valid.

\begin{figure}
    \centering
    \includegraphics[scale=0.5]{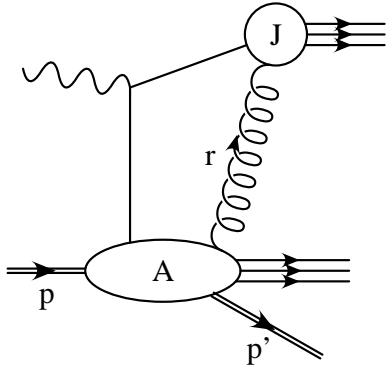}
\caption{Soft gluon exchange between proton subgraph and jet
    subgraph.}
\label{fig:one.soft.gluon}
\end{figure}

\begin{figure}
    \begin{center}
        \includegraphics[scale=0.6]{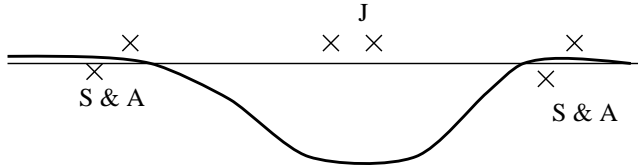}
    \end{center}
\caption{Contour deformation for gluon momentum $r^{+}$ in Fig.\
        \protect\ref{fig:one.soft.gluon}.  The poles are
        labeled by `A', `S' or `J' according to the subgraph
        that causes them.
}
\label{fig:contour}
\end{figure}

Instead, we appeal to a deformation of the other longitudinal
momentum component $r^{+}$.  The simplest case is the exchange of a
single soft gluon, Fig.\ \ref{fig:one.soft.gluon}.  We already
know that to obtain the soft approximation where this gluon
attaches to the {\em final-state} jet subgraph, we must deform
the $r^{+}$
contour away from the (final-state) poles in the jet subgraph.
The limits to this deformation are when the contour reaches the
pole in the putative soft gluon propagator at $r^{+} = r_{T}^{2}/2r^{-}$ or
one of the poles in the $A$ subgraph at $r^{+} \sim Q$:
Fig.\ \ref{fig:contour}.  In either case the contour is deformed
to a region where $|r^{+}| \gg |r^{-}|$, which is not part of the soft
region.  This is sufficient to show that there is no pinch in the
soft region at small $r^{-}$.  Hence we can use the soft
approximation at both the $A$ and $J$ ends of the soft gluon.

Notice that it is not necessary to specify the sign of an $i\epsilon $
for the $1/r^{-}$ factor in Eq.\ (\ref{soft.approx.A}).  Once
subtractions are made to define the soft factor unambiguously, to
remove the collinear contributions, our proof implies that the
soft factor is zero at $r^{-}=0$, and thus the $1/r^{-}$ pole is
cancelled.

\subsection{General soft attachment to $A$}

For the most general case of soft gluons attaching to the
subgraph $A$, we refer back to Fig.\ \ref{fig:regions}.  To get
the desired result we must show that if $r^{-}$ is very small, then
we can deform the $r^{+}$ contour to another region.  This is a bit
tricky, since the deformation may be restricted by poles in other
parts of the soft subgraph, and these give restrictions that are
more severe than those imposed by the poles in the jet subgraphs.

\begin{figure}
    \begin{center}
        \includegraphics[scale=0.5]{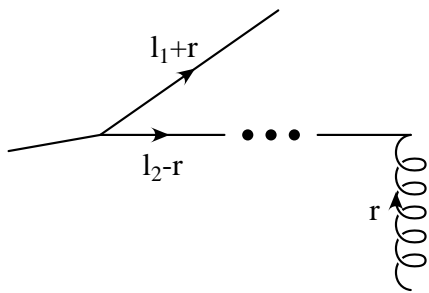}
    \end{center}
\caption{A situation giving a pinch of $r^{+}$.}
\label{fig:obstruction}
\end{figure}

\begin{figure}
    \begin{center}
        \includegraphics[scale=0.5]{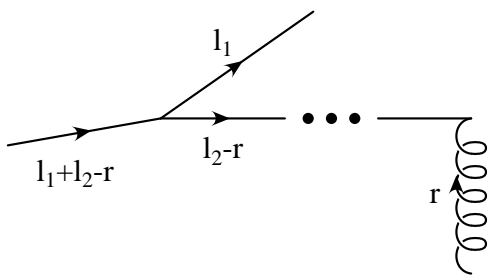}
    \end{center}
\caption{Rerouting $r$ in this way avoids the pinch given by
         Fig.\ \ref{fig:obstruction}.  }
\label{fig:obstruction.avoided}
\end{figure}

The first point to notice is that by hypothesis we start in a
part of the soft region where $r_{T} \gg r^{-}$, the part where Eq.\
(\ref{soft.approx.A}) fails.  This implies that the pole of the
propagator for the line $r$ does not restrict the deformation.

Moreover, the $r^{+}$ contour is {\em not} trapped by the $A$
subgraph. So any pinch would arise from a pinch by other soft
lines or by jet lines.  It would occur only in a situation like
Fig.\ \ref{fig:obstruction}, where we suppose that the lines
$l_{1}+r$ and $l_{2}-r$ both have positive $-$ components of momenta.
Moreover, $l_{1}^{+}$ and $l_{2}^{+}$ must not be much larger than $r^{+}$ and
$l_{1}^{-}$ and $l_{2}^{-}$ must not be so small that the lines are in the
Coulomb region.

But if we do have such a pinch, then we can reroute the momentum
as in Fig.\ \ref{fig:obstruction.avoided}, unless the left-hand
line $l_{1}+l_{2}$ is an external momentum.

So we now have a prescription for avoiding a pinch, if it is
possible at all.  This is to start at the top end of the line
$r$, and to route $r$ {\em back} against the flow of $-$
momentum, as in Fig.\ \ref{fig:obstruction.avoided}.  If by this
procedure we do not arrive at the bottom end of the line $r$,
then we arrive at one of the two incoming lines, either the
proton or the virtual photon.  In either case we can finish the
construction of the route for $r$ by taking it on lines in the
$A$ subgraph. Since by definition these have large $+$ momenta,
while $r^{+}$ is small, none of these lines contribute to a possible
pinch of $r^{+}$.

This completes the proof that the contour of integration over
loop momenta is not trapped in a region where the soft
approximation Eq.\ (\ref{soft.approx.A}) fails for the attachment
of a soft gluon to the $A$ subgraph.

\section{Non-perturbative final-state interactions}
\label{sec:non-pert}

The above proof of factorization relies strictly on the power
counting obtained in perturbation theory.  We now show that
non-perturbative soft effects do not affect the proof, at least
in the context of normal models, such as those appropriate to the
soft pomeron physics treated in Refs.\ \cite{CW,DEL} for the case
of the Drell-Yan process.

One of the key points that enabled us to use the soft
approximation, Eq.\ (\ref{soft.approx}), was that in finite order
perturbation theory the only soft subgraphs that give a leading
power are those which attach to the collinear subgraphs purely by
gluon lines.  Any such soft gluon joins two vertices with momenta
of very different rapidities, so that the vertex $\Gamma ^{\mu }$ in Eq.\
(\ref{soft.approx}) can be replaced by $\Gamma ^{-}n_{J}^{\mu }$.

We know that there must be non-perturbative final-state
interactions that perform hadronization, and that these
interactions give a distribution of particles with several per
unit rapidity.  These interactions can be represented by graphs
like Fig.\ \ref{fig:regions} except that the soft attachments to
the jets are not purely gluons joining vertices of very different
rapidities.  In a perturbative model of this situation, to get a
contribution that does not fall off as a power of $Q$, the
rapidities carried by lines in the graph must cover the whole
range from the rapidity of $A$ to the rapidity of $J$, without
large gaps.  This implies that the order of the graph must be at
least of order the available rapidity range, i.e., the order of
the relevant graphs increases at least as fast as $\ln Q$ at
large $Q$.

Luckily, the second part of the argument leading to the soft
approximation still applies, that is, the contour deformation.
In general, when the momentum transfer $s^{\mu }$ across the subgraph
$S$ is associated the non-perturbative hadronization
interactions, we expect $s^{\mu }$ to have components of order
$(\Lambda /Q^{2},\Lambda /Q^{2},\Lambda )$.  Once we deform $s^{+}$ to values
of order $\Lambda $ or
bigger, as is the result of our argument, the jet lines in which
$s^{\mu }$ flows become off-shell by order $\Lambda Q$.  We now have a
perturbative region where we can use the usual power-counting
rules.

This argument is very similar to arguments used before the advent
of QCD to prove that parton model formulae are valid.  See, for
example, Refs. \cite{CW,DEL,LP}.  In those arguments it was
assumed that the result of contour deformations such as we
perform is that the contours can be taken to infinity with a zero
result --- the assumption of soft behavior of vertices.  In QCD
we cannot take the contours to infinity, but instead we take the
contours from the original region to one that we can treat either
purely perturbatively or with the aid of Ward identities.

\section{Conclusions}
\label{sec:concl}

We have proved the factorization theorem for the general class of
diffractive deep-inelastic processes, and generalizations
including those to which the fracture function formalism of
Trentadue and Veneziano \cite{TV} applies.  The proof includes a
treatment of non-perturbative effects at the level of Refs.\
\cite{CW,DEL,LP}.

Given the results of Refs.\ \cite{CW,DEL} on the Drell-Yan
process, we must not expect the theorem to be applicable to
hadron-hadron collisions. Absorptive corrections should reduce
diffractive hard-scattering cross sections below the expectations
given by the factorization formula on the basis of deep-inelastic
data. Furthermore, the `coherent pomeron mechanism' of
\cite{CFS,BS1,BC} may exist. It is only when one of the
initiating particles is a lepton that the proof of factorization
is valid.

The proof would appear to apply also to {\em direct}
photo-production of jets, etc., because the initiating particle
of the hard scattering is a lepton.  However, the proof does not
apply to {\em resolved} photoproduction processes, since these
are in effect hadron-hadron processes.  The lack of an absolutely
unambiguous separation between direct and resolved
photoproduction will presumably limit the accuracy of the
application of the factorization formula to direct
diffractive photoproduction.

{\em Note added:} After completion of this paper, a paper by
Grazzini, Trentadue, and Veneziano \cite{GTV} appeared, in which
the concept of an `extended fracture function' is defined, with
the aid of the cut-vertex formalism of Mueller \cite{cut-vertex}.
Extended fracture functions are exactly the same as the
diffractive parton densities I define in this paper; they are
fracture functions without the integral over the transverse
momentum of the detected final-state hadron.  Grazzini et al.\
give a brief proof of factorization in the case of $(\phi ^{3})_{6}$
theory. This theory is simpler than QCD since soft exchanges are
power suppressed.  Given this fact, the proofs and results in the
paper of Grazzini et al.\ are completely compatible with those in
the present paper.

\section*{Acknowledgments}

This work was supported in part by the U.S.\ Department of Energy
under grant number DE-FG02-90ER-40577.
I would like to thank CERN and DESY for their support and
hospitality while this work was performed, and
I would like to thank W. Buchm\"uller, D. de Florian, L.
Frankfurt, D. Graudenz, A. Hebecker, P. Landshoff, D. Soper, G.
Sterman, M. Strikman, T. Teubner, and G. Veneziano for
conversations.


\end{document}